\documentclass[%
 aip,apl,
 amsmath,amssymb,
 reprint,%
]{revtex4-1}

\makeatletter

\usepackage{graphicx}
\usepackage{dcolumn}
\usepackage{bm}
\usepackage{xr}
\usepackage{xcolor}
\usepackage{ulem}
\usepackage[utf8]{inputenc}
\usepackage[T1]{fontenc}
\usepackage{mathptmx}
\usepackage{ulem}
\usepackage{multirow}
\usepackage{tabu}
\usepackage{xr}
\usepackage{makecell}

\newcommand{\ec}{E_{\rm{C}}}

\begin{document}

\preprint{AIP/123-QED}

\title{Dispersive sensing in hybrid InAs/Al nanowires}

\author{Deividas~Sabonis}
\thanks{These authors contributed equally to this work}
\affiliation{Microsoft Quantum Lab Copenhagen and Center for Quantum Devices, Niels Bohr Institute, University of Copenhagen, Universitetsparken 5, 2100 Copenhagen, Denmark}

\author{Eoin~C.T.~O'Farrell}
\thanks{These authors contributed equally to this work}
\affiliation{Microsoft Quantum Lab Copenhagen and Center for Quantum Devices, Niels Bohr Institute, University of Copenhagen, Universitetsparken 5, 2100 Copenhagen, Denmark}

\author{Davydas~Razmadze}
\affiliation{Microsoft Quantum Lab Copenhagen and Center for Quantum Devices, Niels Bohr Institute, University of Copenhagen, Universitetsparken 5, 2100 Copenhagen, Denmark}

\author{David~M.T.~van~Zanten}
\affiliation{Microsoft Quantum Lab Copenhagen and Center for Quantum Devices, Niels Bohr Institute, University of Copenhagen, Universitetsparken 5, 2100 Copenhagen, Denmark}

\author{Judith~Suter}
\affiliation{Microsoft Quantum Lab Copenhagen and Center for Quantum Devices, Niels Bohr Institute, University of Copenhagen, Universitetsparken 5, 2100 Copenhagen, Denmark}

\author{Peter~Krogstrup}
\affiliation{Microsoft Quantum Materials Lab and Center for Quantum Devices, Niels Bohr Institute, University of Copenhagen, Kanalvej 7, 2800 Kongens Lyngby, Denmark}

\author{Charles~M.~Marcus}
\affiliation{Microsoft Quantum Lab Copenhagen and Center for Quantum Devices, Niels Bohr Institute, University of Copenhagen, Universitetsparken 5, 2100 Copenhagen, Denmark}

\begin{abstract}
Dispersive charge sensing is realized in hybrid semiconductor-superconductor nanowires in gate-defined single- and double-island device geometries. Signal-to-noise ratios (SNRs) were measured both in the frequency and time domain. Frequency-domain measurements were carried out as a function of frequency and power and yield a charge sensitivity of ~1~$\cdot$~$10^{-3}$~e/~$\sqrt[]{\text{Hz}}$ for an $\sim$11~MHz measurement bandwidth. Time-domain measurements yield ${\rm SNR}>1$ for $20~\mu$s integration time. At zero magnetic field, photon-assisted tunneling was detected dispersively in a double-island geometry, indicating coherent hybridization of the two superconducting islands. At an axial magnetic field of 0.6 T, subgap states are detected dispersively, demonstrating the suitability of the method for sensing in the topological regime.
\end{abstract}
\maketitle

Readout of quantum systems on timescales short compared to coherence or relaxation times is typically performed by one of a few  schemes: (i) the device is incorporated into a resonant circuit, allowing state-dependent changes in the damping or shift of the resonance to be measured, \cite{Schoelkopf1998,bar} (ii) the quantum state is converted to charge, which is then detected by a nearby electrometer,\cite{doi:10.1063/1.4929827,PhysRevApplied.5.034011,PhysRevApplied.8.054006,doi:10.1063/1.2794995} or (iii) a state-dependent capacitive coupling to the system results in a frequency shift in the coupled resonant circuit that depends on the quantum state,\cite{PhysRevLett.95.206807,PhysRevLett.110.046805} the latter referred to as dispersive readout.  In the context of topological qubits, several proposals for non-locally encoding fermion parity in Majorana zero modes have been made\cite{sarma_majorana_2015,Aasen2016,Karzig2017,vonOppen,LiangFu}. Some proposals use parity-to-charge conversion for readout \cite{Aasen2016}, while others use state-dependent hybridization of the Majorana mode with an ancillary system, leading to a dispersive readout signal.\cite{Karzig2017,vonOppen}

Integrating readout circuitry into an existing electrostatic gate or ohmic contact is useful for reducing device footprint and lead count.\cite{PhysRevLett.110.046805,west_gate-based_2019,
doi:10.1063/1.4962811,Zheng,PhysRevApplied.10.014018,doi:10.1063/1.4984224,Urd,PhysRevApplied.9.054016, hornibrook2014frequency, gonzalez-zalba_probing_2015,doi:10.1021/acs.nanolett.5b01306,doi:10.1021/acs.nanolett.6b04354,doi:10.1021/acs.nanolett.5b04356,doi:10.1021/acs.nanolett.5b04356} In this case, dispersive readout is performed by monitoring state-dependent shifts in the resonance frequency $f_{\rm R} = (LC_{\rm tot})^{-1/2}$ of an $LC$ circuit connected to a gate, where $f_{\rm R}$ is  detuned from the qubit transition frequency. The total capacitance, $C_{\rm tot}$, comprises geometric capacitance, $C_{\rm g}$ (including parasitic contributions), quantum capacitance, $C_{\rm Q}$, and tunnel capacitance, $C_{\rm T}$. \cite{PhysRevLett.95.206807,PhysRevB.95.045414} When the quantum system consists of a Coulomb island tunnel coupled to a reservoir, $C_{\rm Q}$ arises from continuous charge transitions, and is proportional to the curvature of energy with respect to the confining gate voltage.\cite{doi:10.1021/nl100663w} The maximum magnitude of $C_{\rm Q}$ occurs at gate voltages corresponding to charge degeneracy, with opposite signs for ground and first excited states. $C_{\rm T}$ is significant when the energy relaxation rate exceeds $f_{\rm R}$. The dependence of $f_{\rm R}$ on $C_{\rm Q}$ provides the quantum state selectivity of the dispersive shift. Monitoring phase or magnitude of the signal reflected from the resonant circuit thus allows readout of the quantum state of the system. 

\begin{figure*}
    \includegraphics[width=0.97\textwidth]{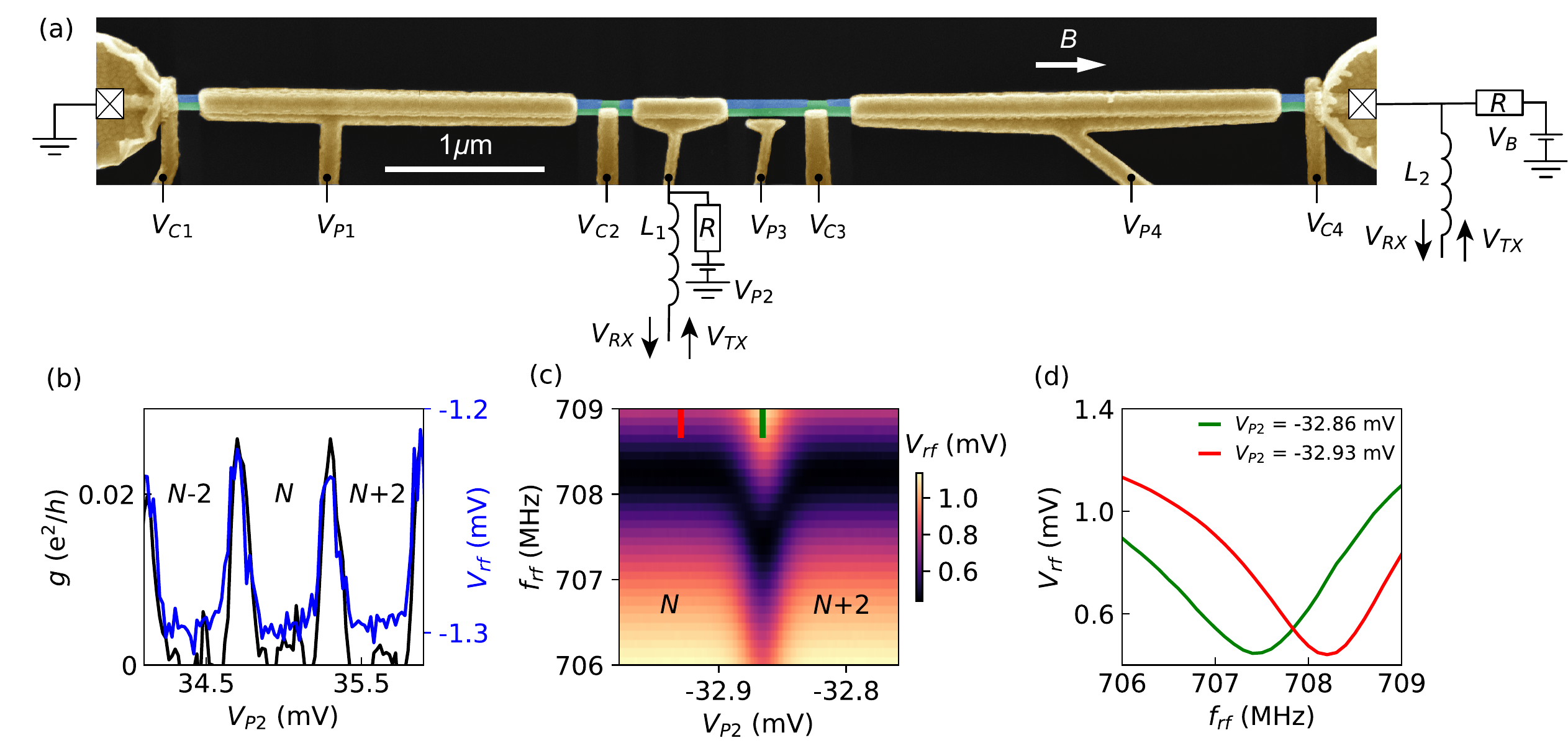}
    \caption{(a) Scanning electron micrograph of the nanowire device together with the relevant electrostatic gates labeled. (b) Coulomb blockade oscillations of superconducting single-island defined between gates C2 and C3 as a function of plunger gate voltage $V_{\rm P2}$ recorded in conductance $g$ via lock-in transport (black) and dispersive lead sensing $V_{\rm rf}$ (blue) showing close matching. (c) Lead sensor signal $V_{\rm rf}$ as a function of electrostatic plunger gate voltage $V_{\rm P2}$ tuning the superconducting island over the Coulomb degeneracy as a function of readout frequency $f_{\rm rf}$. (d) Line cuts from (c) on- (green) and off- (red) Coulomb degeneracy. Dispersive shift in frequency can be observed as the island is tuned over the degeneracy.}
    \label{fig:1}
\end{figure*}

Recent work on gate-based dispersive sensing has addressed semiconducting nanowires (NWs) \cite{doi:10.1021/nl501242b,damaz} and semiconductor quantum dots coupled to subgap states in semiconductor-superconductor NWs at zero magnetic field. \cite{vanVeen}
Beyond qubit readout, dispersive sensing has been used to allow rapid tuning of quantum devices, yielding complementary information to conventional transport approaches.\cite{jarratt2019dispersive} 

In this Letter, we explore dispersive charge sensing in an epitaxial semiconductor-superconductor (InAs/Al)  nanowire device configured to form either a single or double island depending on gate voltages. Since both proximity-induced superconductivity and high magnetic fields are needed for realizing the topological regime, we have focussed particular attention on operation at magnetic fields compatible with topological superconductivity. At zero field, we extract a signal-to-noise ratio (SNR) of a gate-based dispersive sensor as a measure of its sensitivity. Time-domain measurements gave ${\rm SNR} > 1$ for integration times $> 20~\mu$s, as described in detail below. Applying a continuous microwave drive to a nearby gate induces photon assisted tunneling (PAT) \cite{PhysRevLett.93.186802,van2019photon}, indicating coherent hybridization of the two islands.

The device, shown in Fig.~\ref{fig:1}(a), is based on an InAs/Al NW with 7 nm of epitaxial Al grown on three facets of the hexagonal cross section. \cite{Krogstrup2015} Following deposition of individual wires on a Si-SiOx substrate, three 100 nm segments of Al were removed by wet etching to provide tunable tunnel barriers. An insulating layer of HfO$_2$ was then deposited over the wire and electrostatic gates labeled C1, C2, C3 and C4 were deposited, creating three segments of lengths $\sim$~2.5$\,\mu$m,~1$\,\mu$m, and~3.5$\,\mu$m, separated by gate-voltage-controlled barriers of InAs only (see Ref.~\onlinecite{Supplement} for fabrication details). The detector gate, labeled P2, and the right ohmic contact were bonded to superconducting spiral inductors \cite{hornibrook2014frequency} fabricated on a separate sapphire chip to form a resonant circuit that was capacitively coupled to a conventional radio-frequency (rf) reflectometry detection chain. Each resonator was also connected to bias resistors, allowing DC voltages to be applied to the gates. Data from two devices (D1 and D2) are presented. All measurements were made in a dilution refrigerator with base temperature $\sim 20$~mK equipped with a vector magnet (see Ref.~\onlinecite{Supplement} for measurement details).

We initially consider zero magnetic field. Electrostatic gates C2 and C3 were set to the tunneling regime, forming a superconducting single island. A comparison of conductance measured via low-frequency transport using a lock-in amplifier versus reflectometry from the right ohmic contact is shown in Fig.~\ref{fig:1}(b). The conductance, $g$, around zero bias showed Coulomb blockade (CB) peaks as a function of plunger voltage $V_{\rm P2}$. Finite bias conductance measurements yielded a charge energy of $\ec = 60\,\,\mu$eV. CB oscillations had a period of two electron ($2e$) charge. This is expected for $\Delta> \ec $, where $\Delta\sim 180 \,\,\mu$eV is the induced superconducting gap, and indicates low average quasiparticle poisoning. Figure~\ref{fig:1}(c) shows the ohmic reflectometer response as a function of plunger voltage $V_{\rm P2}$ and rf readout frequency $f_{\rm rf}$, see Ref.~\onlinecite{Supplement} for details. The resonance shows a state-dependent shift when crossing the island degeneracy (Fig.~\ref{fig:1}(d)). As shown later, PAT was observed at a frequencies $\geq11$ GHz, i.e., highly detuned from the readout resonators, indicating a state-dependent dispersive interaction between the resonator and the device.

In a lithographically similar device (D2), the gate sensor sensitivity at zero magnetic field was evaluated in a superconducting single-island regime, the island had charging energy $\ec=105\,\,\mu$eV. A nearby gate (P3 in Fig.~\ref{fig:1}(a)) was modulated with a sinusoidal signal of fixed frequency ($f_{\rm M}$) and amplitude $V_{\rm M}$. \cite{doi:10.1063/1.2794995,PhysRevLett.110.046805,PhysRevApplied.9.054016}. Positioning the island gate $V_{\rm P2}$ on the side of a CB peak amplitude modulates the island charge, and thereby the readout resonance, inducing sidebands that are symmetrically detuned by $f_{\rm M}$ from the carrier frequency, $f_{\rm rf}$. 

\begin{figure}
\includegraphics[width=0.48\textwidth]{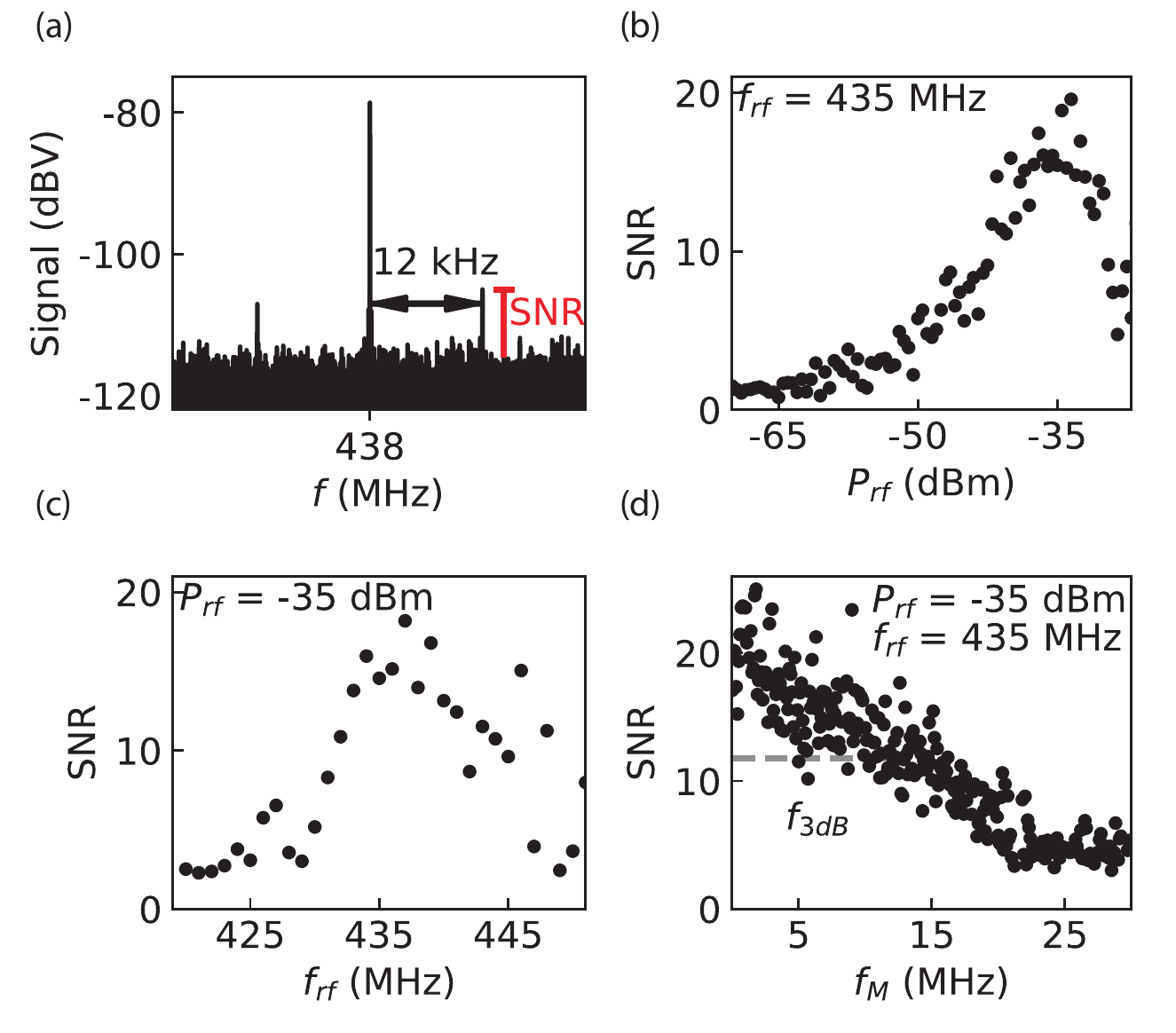}
\caption{\label{fig:SNR1} (a) Spectrum reflected from the detector gate P2 under gate-modulation as described in the main text, two sidebands symmetrically detuned from resonance were observed. (b) Signal-to-noise ratio dependence on carrier power $P_{\rm rf}$ and (c) on carrier frequency $f_{\rm rf}$. (d) Signal-to-noise ratio as a function of modulation frequency $f_{\rm M}$. The 3 dB frequency, $\sim 11\,$MHz, indicates the bandwidth of the measurement.}
\end{figure}

Figure~\ref{fig:SNR1}(a) shows the signal reflected from the detector gate P2 recorded with a spectrum analyser with $\Delta f$~=~13.4~Hz resolution bandwidth. Around the resonance frequency ($\sim$~438~MHz) two sidebands are observed at $f_{\rm M}=\pm12$~kHz. For the analysis that follows the upper sideband was chosen. Signal-to-noise ratio (SNR) is given by the ratio of the height of the sideband to the noise floor in a given bandwidth. The SNR dependence on the rf carrier power $P_{\rm rf}$ (before $\sim$40~dB attenuation) is shown in Fig.~\ref{fig:SNR1}(b). An initial increase in SNR up to around $P_{\rm rf}  = -35$~dBm is observed followed by a decrease for larger power. This turnaround behavior can be explained by the tradeoff between lifetime and power-induced broadening. \cite{PhysRevApplied.10.014018} Figure~\ref{fig:SNR1}(c) shows the SNR dependence on the $f_{\rm rf}$ for a fixed $P_{\rm rf} =-35$~dBm. A maximum SNR was observed at $f_{\rm rf}$~$\sim$~435~MHz. The full-width half-maximum of the SNR as a function of $f_{\rm rf}$ indicates an approximate resonator bandwidth of $\sim 12.2$~MHz.

With $P_{\rm rf}$ and $f_{\rm rf}$ set to maximize SNR [see Figs.~\ref{fig:SNR1}(b, c)], a detection bandwidth of $\sim11\,$MHz was determined~\cite{PhysRevLett.110.046805} by measuring the modulation frequency $f_{\rm M}$ at which SNR decreased by 3~dB, as shown in Fig.~\ref{fig:SNR1}(d).  We next evaluate the charge sensitivity $S\equiv\Delta q(2\Delta f)^{-1/2}10^{-{\rm SNR}/20}$, measured in the time domain,\cite{doi:10.1063/1.2388134} taking spectral resolution $\Delta f$~=~13.4~Hz and SNR~$\sim$~15 as a typical value for optimal detection parameters [see Figs.~\ref{fig:SNR1}(b,c)]. The effective charge change induced by modulation at amplitude $V_{\rm M}$ = 30~m$\text{V}_{\text{pp}}$ was $\Delta q= e (30 {\rm m}\text{V}_{\text{pp}}/950{\rm m}\text{V}_{\text{pp}})\sim0.03e$, found by comparing the amplitude of $V_{\rm M}$ (30~m$\text{V}_{\text{pp}}$) to the amplitude needed to sweep over a full CB peak spacing (950~m$\text{V}_{\text{pp}}$). The factor 1/$\sqrt{2}$ accounts for the power collected from both sidebands. The resulting charge sensitivity was $S \sim 1\cdot10^{-3}e/\sqrt[]{\text{Hz}}$.

\begin{figure}
\includegraphics[width=0.48\textwidth]{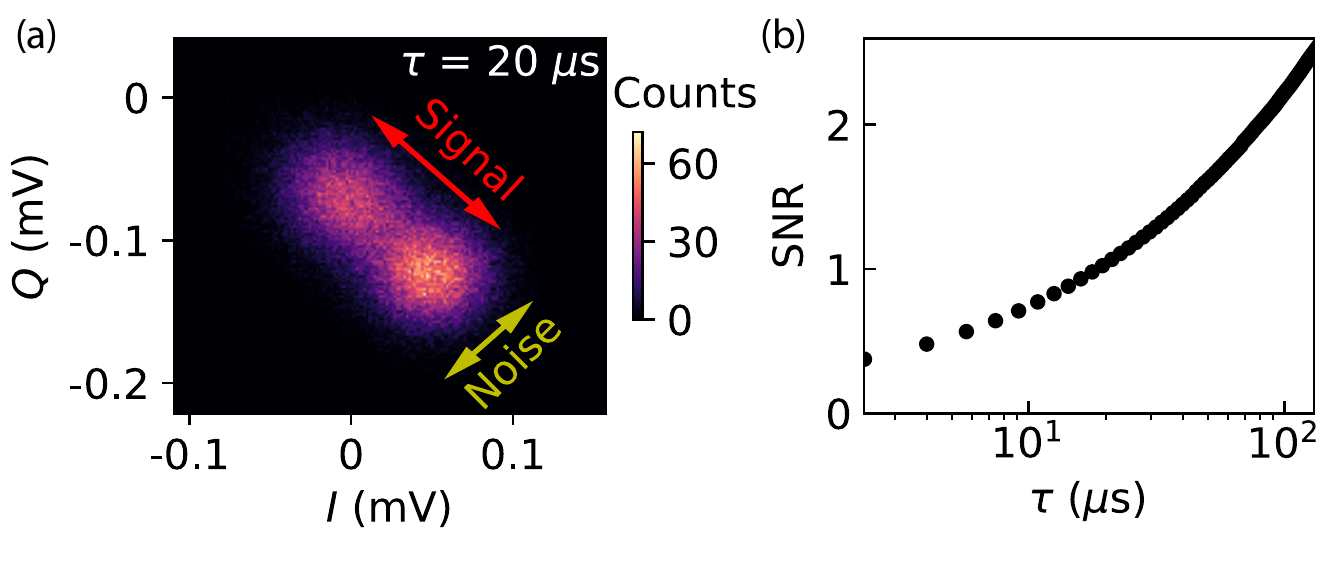}
\caption{(a) Two distinguishable states indicated in in-phase ($I$) and quadrature ($Q$) plane together with "Signal" $\delta V$ defined as a separation between two IQ space maxima and "Noise" $2\sigma$ as induced broadening of the two maxima. (b) Signal-to-noise ratio ($\delta V$/$2\sigma$) as a function of integration time $\tau$.}
\label{fig:SNR2}
\end{figure}

The time needed to obtain a particular SNR at optimal values $f_{\rm rf}$~=~438~MHz and $P_{\rm rf}=-40$~dBm was determined by comparing the difference, $\delta V$, in signal $V_{\rm rf}$ on and off a CB peak with the noise of the measurement, which decreased with increasing signal averaging. In the single-island regime, gate P2 was pulsed on and off a CB peak with amplitude equivalent to $\sim 0.3e$ at a repetition frequency of $2\,$kHz. In-phase ($I$) and quadrature ($Q$) components were recorded with time constant 500~ns, averaged over an integration time $\tau$, and plotted in the complex \textit{I} - \textit{Q} plane. Figure \ref{fig:SNR2}(a) shows an example with $\sim 1.1 \times 10^{4}$ points averaged for $\tau=20\,\mu$s each. SNR$(\tau)$ was found by fitting to two 2D Gaussians, yielding signal $\delta V$ and noise, $2\sigma$. The time-domain SNR, given by $\delta V$/$2\sigma(\tau)$ is shown in Fig.~\ref{fig:SNR2}(b). SNR~$\sim$~1 is reached for integration time $\tau\sim~20\,\mu$s. 

\begin{figure}
    \includegraphics[width=0.5\textwidth]{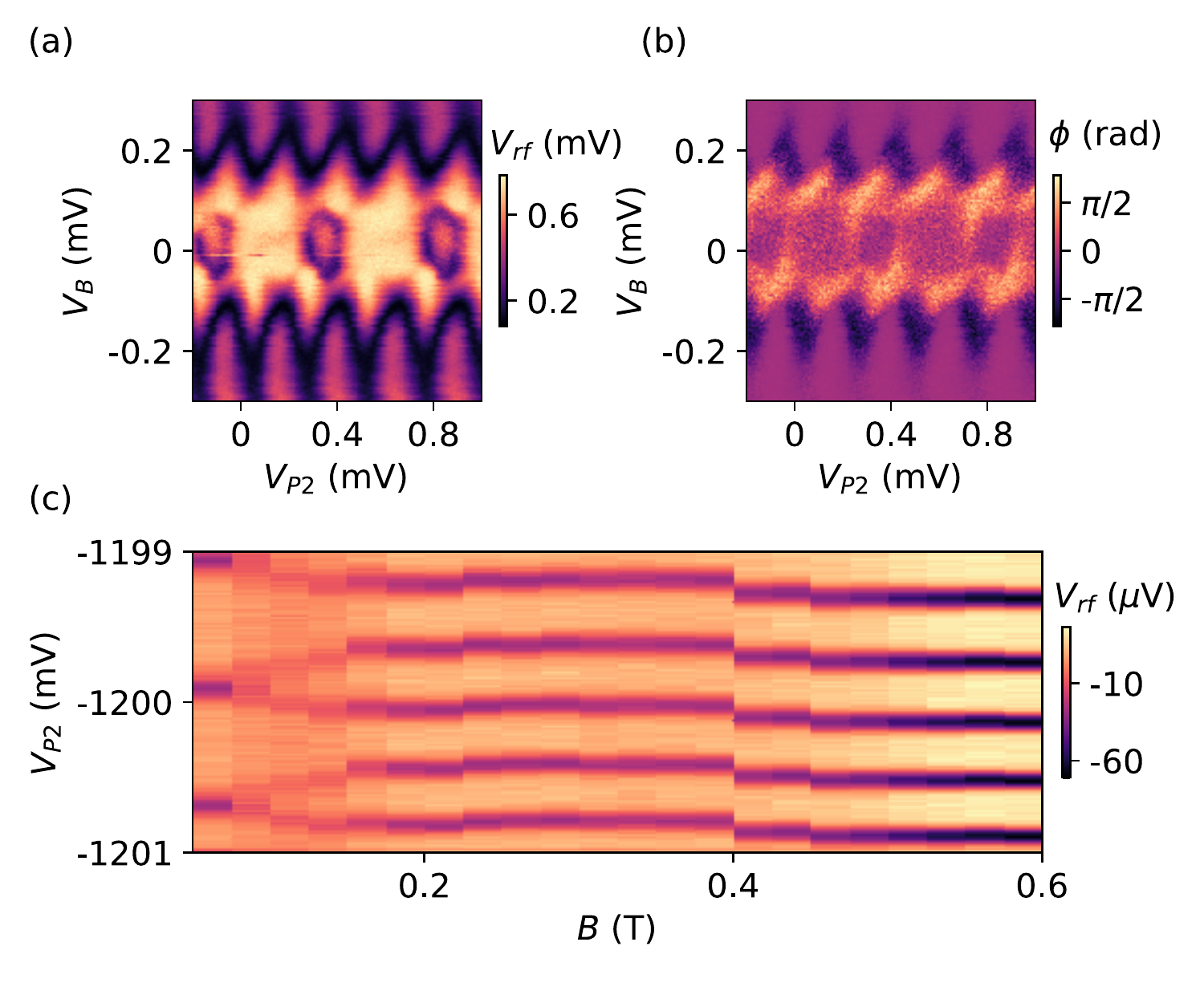}
    \caption{(a) Gate P2 sensing of Coulomb blockaded superconducting single-island recorded in magnitude $V_{\rm rf}$ (a) and phase $\phi$ (b) of the demodulated signal as a function of island plunger voltage $V_{\rm P2}$ and bias $V_{B}$. (c) Coulomb blockade evolution from 2\textit{e} to 1\textit{e} periodic regime as a function of parallel to the nanowire axis magnetic field $B$ and island plunger voltage $V_{\rm P2}$ (line average along $V_{\rm P2}$ axis subtracted).}
    \label{fig:single}
\end{figure}

Figure \ref{fig:single}(a) shows $V_{\rm rf}$ measured using a dispersive sensor on gate P2 [see Fig.~\ref{fig:1}(a)] rather than an ohmic contact. Coulomb diamonds were observed with the device configured as single island at $B=0$. Each measured point in Fig.~\ref{fig:single} was averaged 100 times, yielding a measurement time of $60\,\,\mu$s per point. At high bias, $V_{B}$~$>$~0.2~mV, the period of Coulomb oscillations was halved compared to low bias, indicating that low-bias transport was predominantly carried by Cooper pairs. Figure~\ref{fig:single}(b) shows the phase response of the demodulated signal. The readout frequency was chosen to optimize the $1e$ charge transitions leading to a non-monotonic response for $2e$ charge transitions that had a larger capacitive shift of the resonator. The estimated gate sensor geometric capacitance $C_{\rm g}=e/\delta V_{\rm g}\sim 0.55$~fF, where $\delta V_{\rm g}$ is a gate space periodicity of the island plunger P2. This makes $C_{\rm g}$~$\sim 0.4 C_{\Sigma}$ using the normal-state charging energy of the island.

Application of a parallel magnetic field, $B$, induced subgap states in the nanowire and an evolution from 2\textit{e} to 1\textit{e} periodic CB oscillations, as shown in Fig.~\ref{fig:single}(c), where an average $V_{\rm rf}$ was subtracted at each field. 1$e$ periodic CB oscillations correspond to a state at zero-energy in the superconducting gap,\cite{Albrecht2016} though not necessarily a discrete state. To keep maximum detection contrast at each magnetic field value the readout frequency was adjusted to compensate for changing kinetic inductance of the resonator. The jump at $B$~$\sim$~0.4~T was likely due to electrostatic background charges in the NW environment. The detected signal did not degrade at magnetic field ranges compatible with tuning into a topological state in similar wires.\cite{Deng2016}

\begin{figure}
    \includegraphics[width=0.5\textwidth]{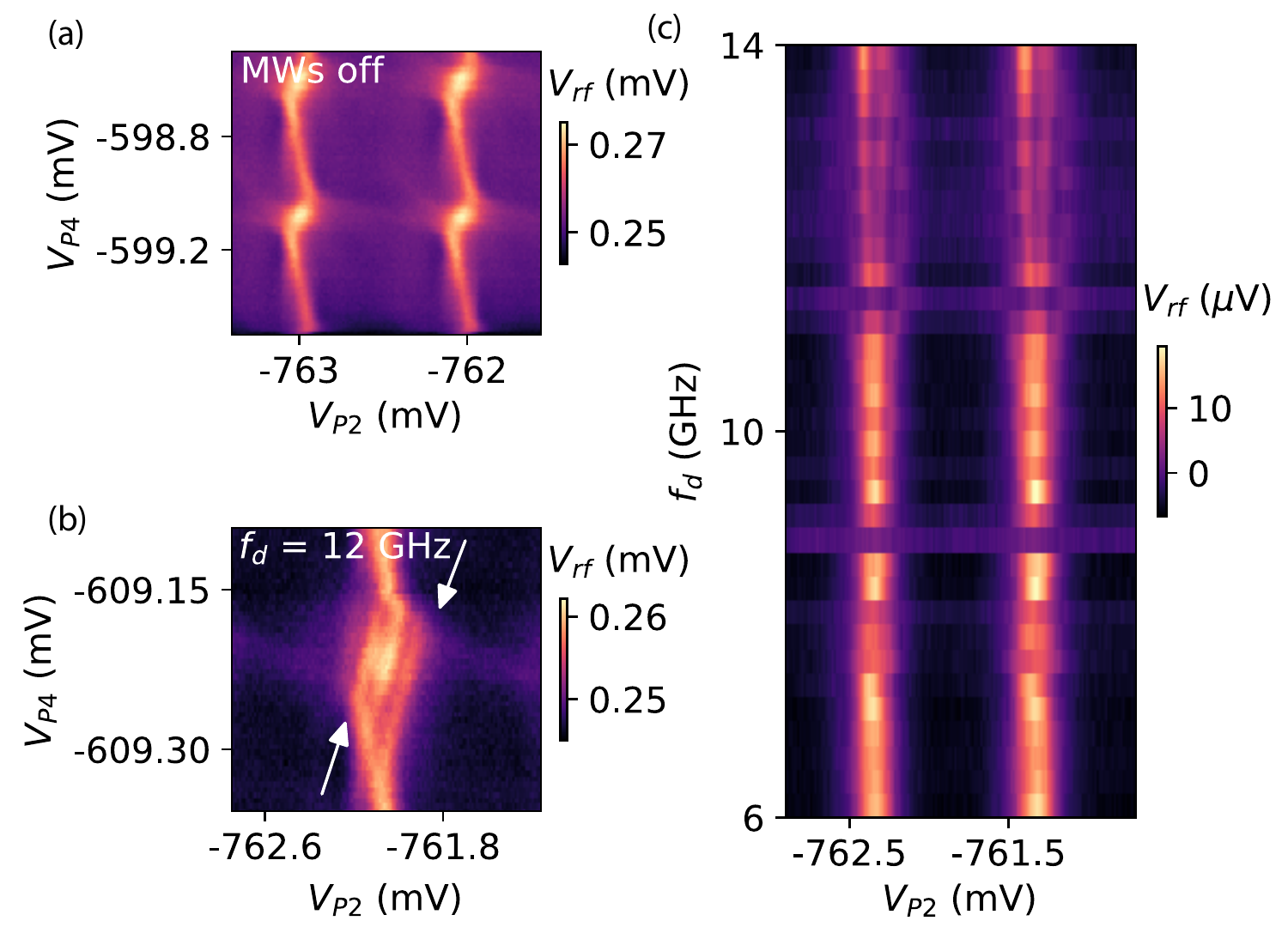}
    \caption{(a) 2\textit{e} periodic charge stability diagram as a function of right ($V_{P4}$) and middle plunger ($V_{\rm P2}$) voltages. (b) Microwave induced photon assisted tunneling transitions between two charge states in a double-island at zero magnetic field. (c) Energy dispersion of the double-island energy levels measured as a function of the plunger voltage $V_{\rm P2}$ and microwave drive frequency $f_{d}$.}
    \label{fig:double}
\end{figure}

The lifetime of the excited state was too short to measure directly, but could be estimated using PAT in a double-dot configuration. At zero field, the 2\textit{e}-periodic double-dot charge stability diagram as a function of voltages $V_{\rm P2}$ and $V_{P4}$ takes on a familiar honeycomb pattern\cite{vdwRMP,Lambert2017,doi:10.1063/1.4729469} as seen in Fig.~\ref{fig:double}(a).  The gate sensor detected two types of transitions: internal, between the two islands, and external, to the left superconducting lead. Transitions to the right normal lead could also be discerned, though with less visibility. Spectroscopy was performed by irradiating the sample with microwaves of frequency $f_{d}$. When $f_{d}$ was comparable to or larger than the tunneling rate of Cooper pairs across the junction, $hf_{d}>E_{J}$, where $E_{J}$ is the Josephson energy, characteristic PAT features were observed in the charge stability diagram\cite{vdwRMP} as seen in Fig.~\ref{fig:double}(b). As usual\cite{PhysRevB.50.2019, Lambert2017, van2019photon}, PAT signatures appear as parallel features along the detuning axis. To quantify the interdot coupling, $f_{d}$ was swept with the system tuned using P4 to cross the interdot charge transitions when gate P2 was also swept. The resulting 2D plot, shown in Fig.~\ref{fig:double}(c) reveals the emergence of PAT features above 11.3~GHz, placing a rough lower bound on $E_{J}$. This extends our previous work on PAT in hybrid double-dot systems \cite{van2019photon} to the technique of dispersive readout. 

In conclusion, gate and lead-based dispersive sensing techniques were applied to Coulomb blockaded single- and double-islands in hybrid semiconductor-superconductor InAs/Al nanowires. Characterization of gate sensing, using sideband modulation, at zero magnetic field, as a function of readout parameters, yielded charge sensitivities of the order of 1~$\cdot$ $10^{-3}$~e/$\sqrt[]{\text{Hz}}$, with a detection bandwidth of $\sim 11$~MHz. In time-domain measurements, SNR of 1 was achieved for an integration time of 20 $\mu$s. Dispersive readout of photon assisted tunneling indicated coherent hybridization of two superconducting islands gave an estimate for the Josephson coupling between islands, $E_{J}$~$\sim$~11.3~GHz. Magnetic field compatibility of the gate sensor up to 0.6~T was demonstrated, compatible with tuning into the topological regime. These results will be useful in employing gate and lead-based dispersive sensing in topological qubit experiments without the requirement of fabricating nearby electrometers.

\vspace{0.5 cm}

\begin{acknowledgments}
We thank Shivendra Upadhyay for help with fabrication, and Ruben Grigoryan, Andrew Higginbotham, John Hornibrook, Ferdinand Kuemmeth, Wolfgang Pfaff, Bernard van Heck and John Watson for useful discussions. Research is supported by Microsoft and the Danish National Research Foundation. PK acknowledges support from the ERC starting grant no.~716655 under the Horizon~2020 framework.
\end{acknowledgments}

\nocite{*}
\bibliography{bibliography}

\setcounter{figure}{0}
\setcounter{equation}{0}
\renewcommand{\theequation}{S.\arabic{equation}}
\renewcommand{\thefigure}{S.\arabic{figure}}
\newpage
\phantom{\ }

\section{Supplementary information for: Dispersive sensing in hybrid InAs/Al nanowires}

\section{Fabrication}
\label{AppA}

Two nanowire devices presented here are around 100~nm in diameter and were grown using the vapor-liquid-solid technique in a molecular beam epitaxy system with the InAs [111] substrate crystal orientation \cite{Krogstrup2015}. Following the NW growth, Al is deposited epitaxially \textit{in situ} on three facets of the NW with an average thickness of 7 nm. The NW is then positioned on a chip with a micro-manipulator tool which gives the ability for few micrometer precision. For both devices the Al was selectively etched using wet etchant Transene D. All patterning was performed using an Elionix ELS-7000 EBL. The InAs/Al NW has Al shell on two of its facets and is fabricated on Si chip covered with 200 nm of $\text{SiO}_{\text{2}}$. The Ti-Au contacts (5 nm + 150 nm) were evaporated after performing rf milling to remove the oxide from the nanowire. Then, 5 nm of $\text{HfO}_{\text{2}}$ was deposited by atomic layer deposition. Finally the last set of Ti-Au gates (5 nm + 150 nm) was evaporated. 
\section{Measurement setup}

\label{AppB}
The resonant circuit, is made out of off-chip superconducting Nb on $\text{Al}_{\text{2}}\text{O}_{\text{3}}$ spiral inductor \cite{hornibrook2014frequency} in series with the distributed parasitic capacitance that includes a \text{Ti}\text{Au} gates. Gates are bonded with 25~$\mu$m diameter Al (1\% Si) bond wires. To reduce parasitic capacitances and maximize sensitivity we tried to minimize the bond wire length. The sample is loaded in a box which is surrounded with Eccosorb microwave absorbent material in order to minimize the effects due to stray radiation. Measurements were performed in an Oxford Instruments Triton 400 dilution refrigerator with a base electron temperature of $T\sim20$~mK and a 1-1-6~T vector magnet. The table \ref{table:1} summarizes the resonant circuit parameters of two measured devices.

\begin{table}[!h]
\centering
  \begin{tabular}{ | c | c | c | c | c | c | c | }
  \hline
   &$L_{1}$ (nH) & $L_{2}$ (nH) & $f_{1}$ (MHz) & $f_{2}$ (MHz) & $Q_{l,\,1}$ & $Q_{l,\,2}$  \\ \hline
   \thead{Device 1 \\ (D1)} & 105 & 150 & 780 & 710 & 95 & 240 \\ \hline      
   \thead{Device 2 \\ (D2)} & 420 & 310 & 440 & 510 & 50 & 60 \\ \hline    
  \end{tabular}
  \caption{Summary of inductor values used for two measured devices together with the bare resonance frequencies and loaded quality factors.}
  \label{table:1}
\end{table}

For reflectometry measurements a commercially available high frequency demodulation unit (Zurich Instruments UHFLI \cite{zi}) was used. The RF carrier (frequency $f_{rf}$, amplitude $V_\mathrm{TX}$) generated at room temperature was sent through high frequency coax line, followed by 21~dB of distributed attenuators with a further 15 dB attenuation from the directional coupler (Minicircuits ZEDC-15-2B) mounted below the mixing chamber plate. The signal reflected from the rf circuit was amplified by approximately +40~dB with 4K amplifier (Caltech CITLF3). The amplified signal (amplitude $V_\mathrm{RX}$) is detected at room temperature using homodyne detection inside of ZI lock-in with phase and magnitude information available for further processing. For microwave spectroscopy measurements in Fig. 4 the signal generator Rohde $\&$ Schwarz, RS SMB100A was used. 

When transport measurements were performed the current $I$ through the NW island was measured by connecting a current amplifier (Low Noise/High stability I/V converter, SP 983 IF3602) to the drain electrode of the device, while applying a voltage bias to the source electrode. Total voltage bias is the sum of a DC ($V_\mathrm{B}$ and a small AC voltage (excitation voltages in the range of 4 - 10~$\mu$V with excitation frequencies below 150~Hz). This allows the measurement of differential conductance, $g\equiv \textup{d}I/\textup{d}V_\mathrm{B}$, by conventional lock-in detection (Stanford Research Systems SR830). The DC gates are connected to twisted pairs with a low pass cut-off frequency of~$\sim$~1~kHz at base temperature. During the rf measurements the DC bias is set to zero unless stated otherwise. We note that the experimental setup resembles the one reported in detail in Ref. \cite{Razmadze2018}. All data was acquired with the modular data acquisition framework QCoDeS \cite{qc}.

\providecommand{\noopsort}[1]{}\providecommand{\singleletter}[1]{#1}%

\end{document}